\documentclass{jors}

\definecolor{mygray}{gray}{0.6}

\usepackage{physics}
\usepackage{amsmath}
\usepackage{amssymb}
\usepackage{bm}
\usepackage[numbers]{natbib}

\usepackage{graphicx}
\graphicspath{{./figures/}}
\usepackage{subcaption}
\usepackage[draft=true]{minted}

\begin{document}

{\bf Software paper for submission to the Journal of Open Research Software} \\

\rule{\textwidth}{1pt}

\section*{(1) Overview}

\vspace{0.5cm}

\section*{Title}

Magpy: A C++ accelerated Python package for simulating magnetic
nanoparticle stochastic dynamics

\section*{Paper Authors}

1. Laslett, Oliver\\
2. Waters, Jonathon\\
3. Fangohr, Hans\\
4. Hovorka, Ondrej

\section*{Paper Author Roles and Affiliations}
1. Engineering and the Environment, University of Southampton,
Southampton, SO17 1BJ, UK.\\
2. Engineering and the Environment, University of Southampton,
Southampton, SO17 1BJ, UK.\\
3. Engineering and the Environment, University of Southampton,
Southampton, SO17 1BJ, UK.
European XFEL GmbH, Holzkoppel 4, 22869 Schenefeld, Germany\\
4. Engineering and the Environment, University of Southampton,
Southampton, SO17 1BJ, UK.\\

\section*{Abstract}


Magpy is a C++ accelerated Python package for modelling and simulating
the magnetic dynamics of nano-sized particles. Nanoparticles are
modelled as a system of three-dimensional macrospins and simulated
with a set of coupled stochastic differential equations (the
Landau-Lifshitz-Gilbert equation), which are solved numerically using
explicit or implicit methods. The results of the simulations may be
used to compute equilibrium states, the dynamic response to external
magnetic fields, and heat dissipation. Magpy is built on a C++
library, which is optimised for serial execution, and exposed through
a Python interface utilising an embarrassingly parallel
strategy. Magpy is free, open-source, and available on github under
the 3-Clause BSD License.

\section*{Keywords}

magnetism; macrospin; nanoparticles; physics; biomedicine;
nanomedicine; stochastic; dynamics; numerical methods; Python; C++

\section*{Introduction}

Magpy is an open-source C++ and Python package that models
nanoparticles and simulates their magnetic state over time. The
magnetic dynamics of atoms within materials are described by the
Landau-Lifshitz-Gilbert (LLG) equation~\cite{lakshmanan2011}, which is
a nonlinear stochastic differential equation (SDE). The best choice of
numerical method to solve the LLG dynamics is an open research
question. Therefore, the numerical solvers in Magpy are implemented in
a generic form, independent from the equations of magnetism, which
allows the methods to be tuned or replaced easily. The current
implementation includes the widely used Heun scheme
\cite{Garcia-Palacios1998} and the first open-source implementation of
the Milstein derivative-free fully implicit scheme
\cite{milstein2002numerical}. Magpy also includes a rare-events model,
valid under additional simplifying assumptions (detailed below), which
avoids solving the LLG equation and consequently simulates single
nanoparticles with significantly less computational effort.

The dynamics of magnetic nanoparticles play a crucial role in a number
of emerging medical technologies. For example, magnetic particles have
been used to enhance MRI images~\cite{na2009inorganic}, deliver drugs
inside the body~\cite{chertok2008iron}, and destroy tumours by means
of heat destruction~\cite{jordan1999}. The particles that
enable these technologies are simulated computationally to augment
traditional experiments and explore the effects of changing material
properties. Although Magpy was designed with medical applications in
mind, the software may be used to explain or predict the outcome of
magnetic nanoparticle experiments in general.

\textbf{A model for magnetic nanoparticle dynamics}

\begin{figure}
  \centering
  \includegraphics[width=\linewidth]{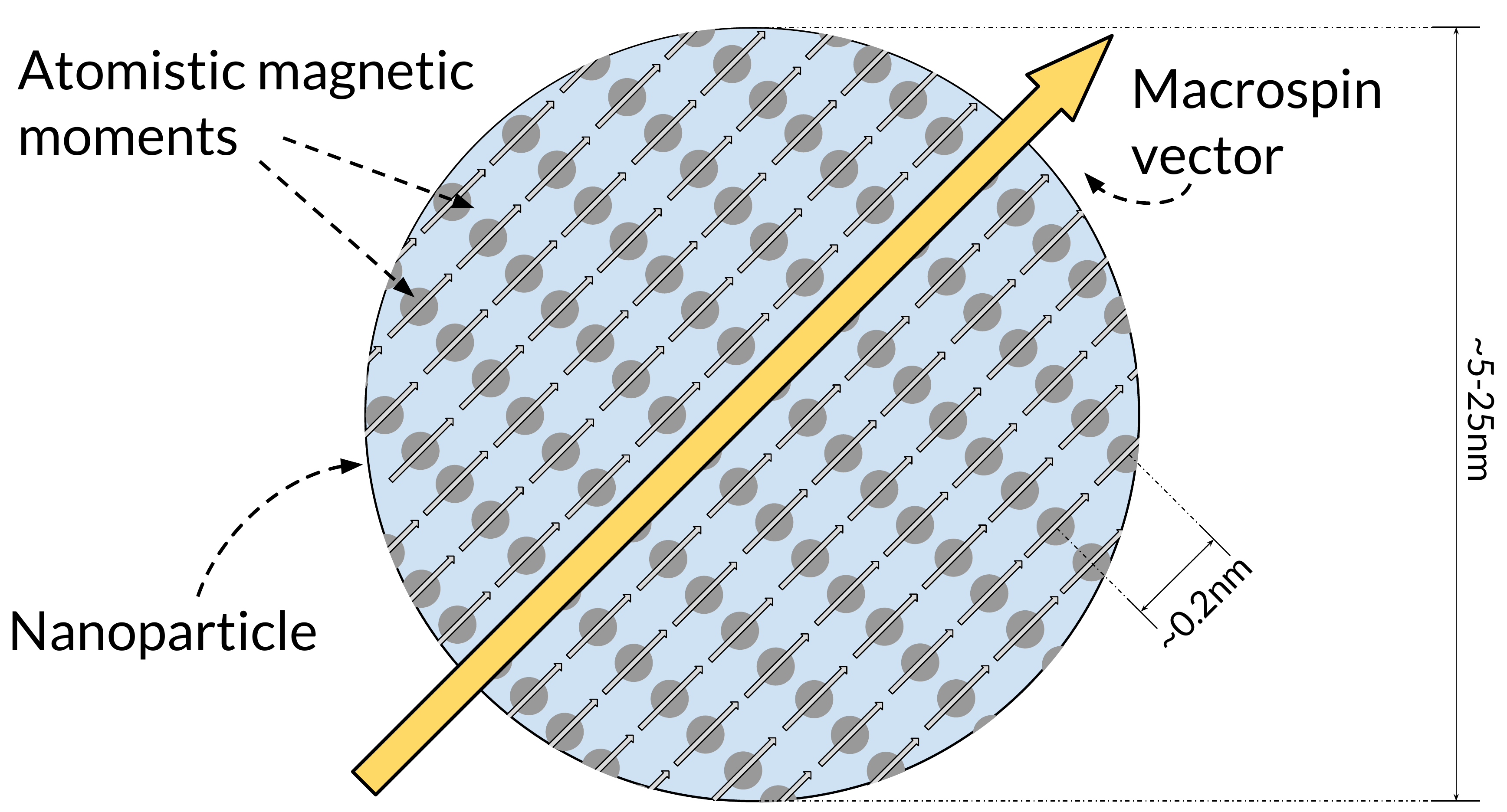}
  \caption{A two-dimensional sketch of a nanoparticle. The atoms of a
    magnetic material are packed into a regular crystal lattice and
    each is modelled by a three-dimensional magnetic moment that
    varies with time. Due to strong interactions between these
    magnetic moments in small particles, Magpy assumes they rotate
    coherently and are represented by a single macrospin
    vector.}
  \label{fig:sketch}
\end{figure}

Figure~\ref{fig:sketch} shows a diagram of a magnetic nanoparticle,
which comprises a large number of individual atoms arranged in a
regular crystal lattice, each of which possesses a magnetic moment
represented by a 3-dimensional vector. The net magnetisation of a
material is simply the sum of the individual magnetic moments. For
example, if the atoms are randomly oriented, the material has zero
magnetisation; if they are all aligned, the material has a large
magnetisation component. Atomic magnetic moments prefer to align with
one another due to an exchange interaction force. This force is strong
enough that, in nanoparticles that are particularly small ($<25$nm in
diameter), the individual moments are approximately aligned and rotate
coherently. Magpy uses this approximation to model the state of a
particles' atoms by a single 3-dimensional vector termed a macrospin,
rather than simulate each atom individually. Magpy is able to simulate
much longer timescales for the same computational effort compared to
simulating the individual atoms due the greatly reduced degrees of
freedom.

\begin{figure}
  \centering
  \includegraphics[width=\linewidth]{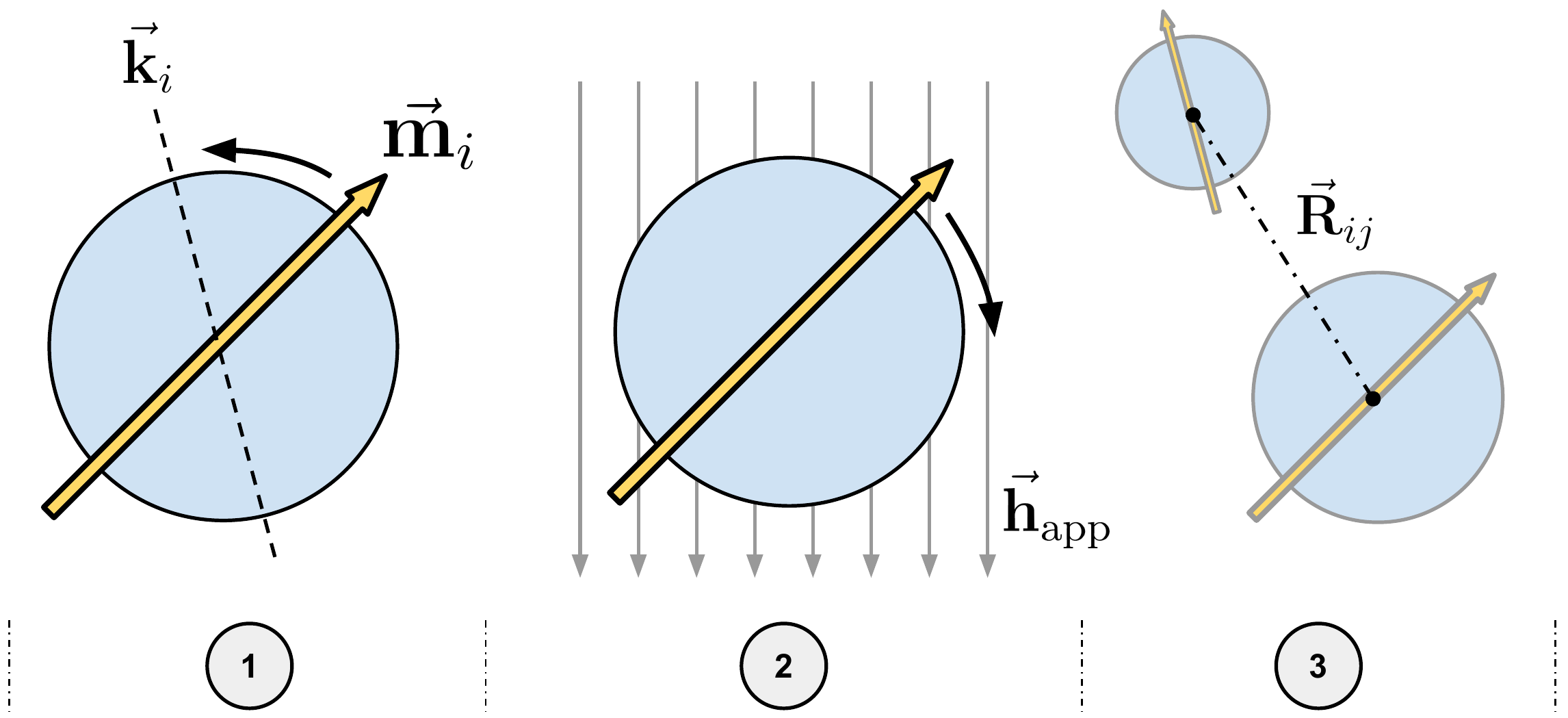}
  \caption{The three effective field contributions acting upon
    macrospin $i$. (1) The macrospin experiences a force towards
    alignment with the particle anisotropy axis (dashed line)
    $\vb{k}_i$ in either direction. (2) The macrospin is also forced
    towards aligning with the externally applied field direction
    (solid arrows) $\vb{h}_\textrm{app}$. (3) Finally, each macrospin
    is repelled and attracted by nearby macrospins. The force of the
    dipole-dipole interaction diminishes with distance (dash-dotted
    line) between two particles $\vb{R}_{ij}$.}
  \label{fig:forces}
\end{figure}

The dynamics of a macrospin are described by the
Landau-Lifshitz-Gilbert equation (LLG). For a detailed explanation of
the origins and derivation of the LLG see~\cite{lakshmanan2011}. The
LLG is implemented in Magpy in a normalised form by replacing time
with a \emph{reduced time} variable $\ell$ (defined below), which
ensures that the variables in the equation are around an order of
magnitude of unity:
\begin{equation}
  \label{eq:llg-norm}
  \dv{\vb{m}_k}{\ell} =
  -\left(\vb{m}_k\times\left(\vb{h}_k + \bm{\xi}_k
    \right)\right) -
  \alpha\vb{m}_k\times\left(\vb{m}_k\times \left(\vb{h}_k +
      \bm{\xi}_k \right)\right)
\end{equation}
where $\vb{m}_k\in\mathbb{R}^3$ is the unit vector of the
$k^{\mathrm{th}}$ particle macrospin, $\alpha$ is a damping constant,
$\bm{\xi}_k\in\mathbb{R}^3$ is a random term that models the thermal
fluctuations experienced by the particle, and $\vb{h}_k$ is the
effective magnetic field experienced by the particle. The effective
field $\vb{h}_i$ comprises three different components
\cite{haase2012role}, which are depicted in Figure~\ref{fig:forces}
and given mathematically as:
\begin{equation}
  \label{eq:heff}
  \begin{split}
    \vb{h}_i=&-k_i\left(\vb{m}_i\cdot\vb{k}_i\right)\vb{k}_i - \vb{h}_\mathrm{app}\\
    &-\sum_{j\neq i} \frac{\mu_0M_\mathrm{s}^2}{2\bar{K}}\frac{
      v_j}{4\pi \left|\vb{r}_{ij}\right|^3}
    \left(3\left(\vb{m}_j\cdot\vb{r}_{ij}\right)\vb{r}_{ij} -
      \vb{m}_j\right)
  \end{split}
\end{equation}
The first term describes preferential alignment of $\vb{m}$ with the
particle's anisotropy axis, which acts in the unit direction
$\vb{k}_i\in\mathbb{R}^3$ with magnitude $k_i$ (units Jm$^{-3}$). The
second term describes the effect of an externally applied field
$\vb{h}_\mathrm{app}\in\mathbb{R}^3$. The final term describes the
field experienced by particle $i$ through a long-range dipole-dipole
interaction with a nearby particle $j$, where $V_j$ is the volume
(units m$^3$) of particle $j$ and $v_j=V_j/\bar{V}$ is the reduced
volume; $\bar{V}=1/N\sum_{n=0}^NV_n$ is the mean volume of all
particles in the system; $\bar{K}=1/N\sum_{n=0}^Nk_n$ is the mean
anisotropy magnitude; $\mu_0=4\pi \times 10^{-7}$ is a constant (units
mkgs$^{-2}$A$^{-2}$); $M_\mathrm{s}$ is the magnitude of the macrospin
(the saturation magnetisation, units Am$^{-1}$) and
$\vb{r}_{ij}\in\mathbb{R}^3$ and $r_{ij}$ are the unit vector and
magnitude respectively of the reduced distance between particles $i$
and $j$. The reduced distance is the true distance (units m) divided
by $\sqrt[3]{\bar{V}}$ and appears in equation (\ref{eq:heff}) because
the numerator and denominator of the interaction term are divided by
$\bar{V}$, which has the effect of scaling both values close to
unity. The prefactor $\mu_0M_\mathrm{s}^2/\qty(2\bar{K})$ can be
computed in advance and will also evaluate close to unity.

The reduced simulation time is related to real time through:
\begin{equation}
  \label{eq:reduced-time}
  \ell = t\frac{2\gamma\bar{K}}{M_\mathrm{s}\qty(1+\alpha^2)}
\end{equation}
where $\gamma=1.76086\times 10^{11}$ is a constant (units
rads$^{-1}$T$^{-1}$). The fluctuating thermal field is a vector of
independent and identically normally distributed random variables
($\xi_k^i\qty(\ell)$ is the $i^{th}$ component of the thermal field
acting on macrospin $k$ at time $\ell$) such that the covariance
between two components is~\cite{Garcia-Palacios1998}:
\begin{equation}
  \label{eq:sllgnoise}
  \left<\xi^i_k\left(\ell\right)\xi^j_k\left(\ell'\right)\right>
  = \delta_{ij}\delta\left(\ell-\ell'\right)
  \sqrt{\frac{\alpha k_BT}{\bar{K}V_k\qty(1+\alpha^2)}}
\end{equation}
where $\delta_{ij}$ and $\delta\qty(\ell-\ell')$ are the Kronecker
delta and Dirac delta functions respectively, $k_B$ is the Boltzmann
constant (units m$^2$kgs$^{-2}$K$^{-1}$) and $T$ is the temperature
(units K).

Equations~\eqref{eq:llg-norm}-\eqref{eq:sllgnoise} describe the Magpy
model of a system of magnetic nanoparticles. The equations are solved
numerically at discrete time steps, resulting in a simulated
trajectory of the system's magnetic state. The simulation outputs, at
each discrete time, the value of the applied magnetic field and the
$x,y,z$ components of the magnetic state of every particle in the
system. These results may be used to obtain the total magnetisation of
the system $M\qty(t)=M_\mathrm{s}\sum_i m_i\qty(t)$, the average magnetisation
of an ensemble of systems, static and dynamic hysteresis loops, and
the energy dissipated by the system.

\begin{figure}
  \centering
  \begin{subfigure}[t]{0.4\textwidth}
    \includegraphics[width=.9\linewidth]{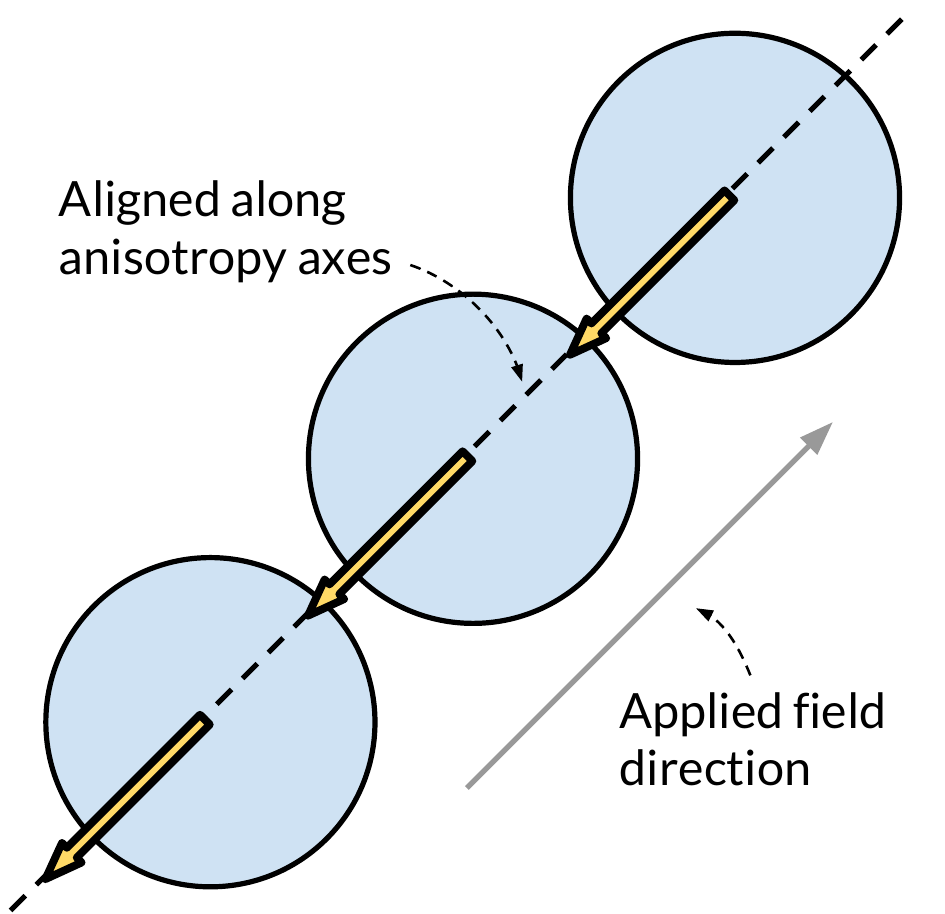}
  \end{subfigure}\hfill%
  \begin{subfigure}[t]{0.6\textwidth}
    \includegraphics[width=\linewidth]{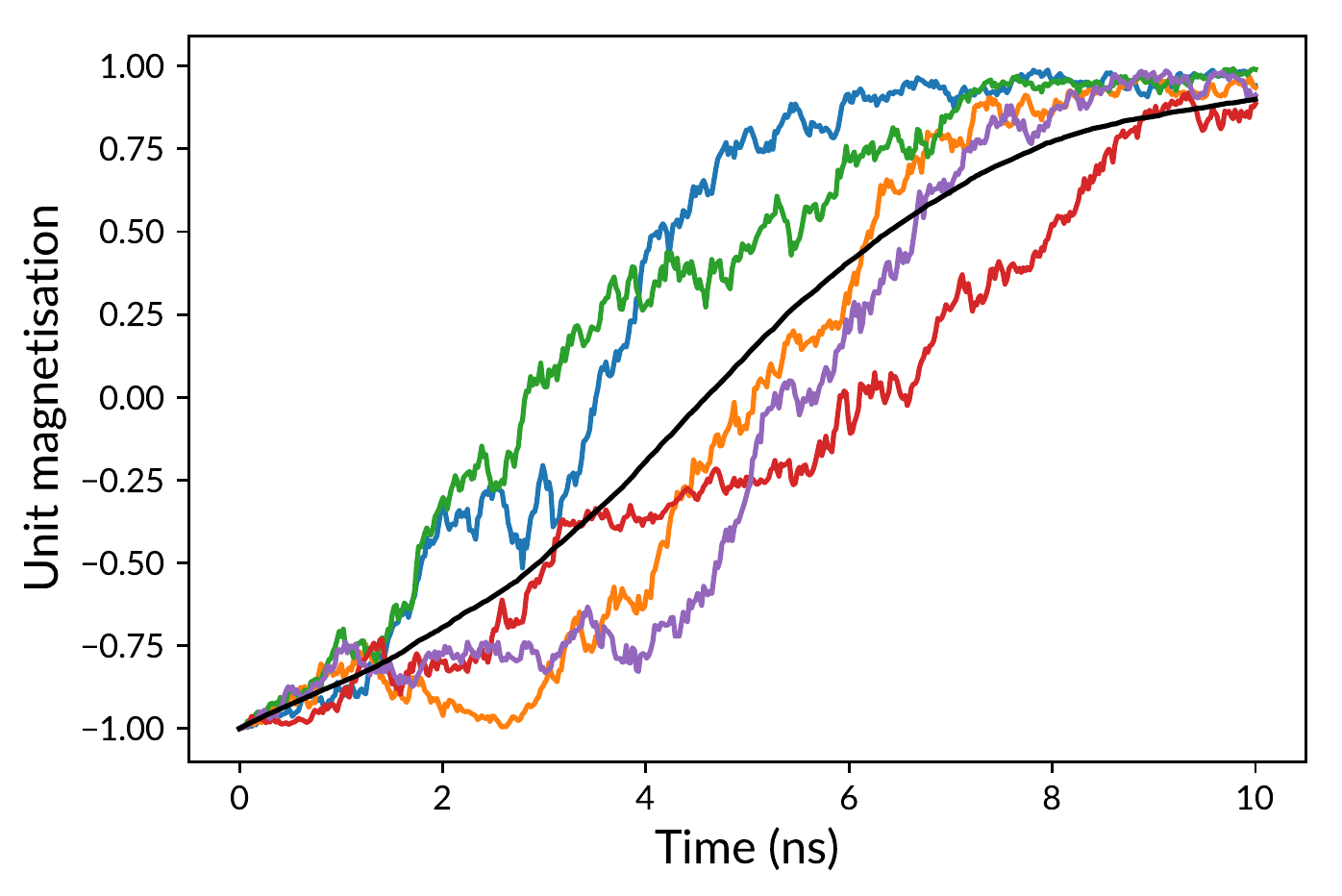}
  \end{subfigure}
  \caption{Simulating a chain of three particles (created in Magpy
    using Listing~\ref{listing:threechain}). (left) A chain structure
    of three identical particles are initialised with an external
    field applied along their anisotropy axis and their magnetisation
    initially against the applied field. (right) The coloured lines
    show the total magnetisation in the direction of the applied field
    for 5 simulations from the same initial condition (see left). The
    black line is the result of averaging 500 simulations from the
    same initial condition (i.e. the expected or mean trajectory).}
  \label{fig:sim-example}
\end{figure}

Multiple simulations with different random seeds but with identical
initial conditions will result in different solutions due to the
stochastic nature of the thermal field. For example
Figure~\ref{fig:sim-example}, shows the results of five simulations of
a 3-particle chain from the same initial condition. The Magpy script
used to generate the results is shown in
Listing~\ref{listing:threechain}. In addition to individual
trajectories, the expected system trajectory and higher order
statistical moments may be obtained by sampling a multitude of
individual simulations (the Monte-Carlo technique).
The number of simulations required to obtain reasonable estimates of
these statistical variables is large and depends on the system of
interest.

\begin{listing}
\begin{minted}
  [
  frame=lines,
  framesep=2mm,
  baselinestretch=1.2,
  bgcolor=white,
  fontsize=\footnotesize,
  linenos
  ]
  {python}
import magpy as mp

chain3_model = mp.Model(
    radius=[8e-9, 8e-9, 8e-9],
    anisotropy=[4e3, 4e3, 4e3],
    anisotropy_axis=[
        [0., 0., 1.], [0., 0., 1.], [0., 0., 1.]],
    magnetisation_direction=[
        [0., 0., -1], [0., 0., -1], [0., 0., -1]],
    location=[
        [0., 0., -20e-9], [0., 0., 0.], [0., 0., 20e-9]],
    magnetisation=400e3,
    damping=0.1,
    temperature=300.,
    field_shape='constant',
    field_amplitude=30e3)

chain3_ensemble = mp.EnsembleModel(base_model=chain3_model, N=500)
results = ensemble.simulate(
    time_step=1e-13, end_time=1e-8, max_samples=500, n_jobs=4)

time = results.time
first_run_magnetisation = results.results[0].magnetisation()
ensemble_magnetisation = results.ensemble_magnetisation()
\end{minted}
\caption{Simulating an ensemble of five hundred three-particle chains
  in Magpy (results shown in Figure~\ref{fig:sim-example}). The
  three-particle chain model is instantiated (lines~3-16) using the
  \texttt{Model} object, which is defined by the properties and
  locations of each particle in the chain and the applied field. An
  ensemble of models is created using the \texttt{EnsembleModel}
  object (line~18), which simply represents a collection of individual
  models and is provided for convenience. The five hundred models are
  individually simulated (lines~19-20) and the computational work is
  distributed across four processes by setting \texttt{n\_jobs=4}. The
  resulting magnetisation is computed for an individual model
  (line~23) and the entire ensemble of models (line~24).}
\label{listing:threechain}
\end{listing}

\begin{figure}
  \centering
  \begin{subfigure}[t]{0.55\textwidth}
    \includegraphics[width=\linewidth]{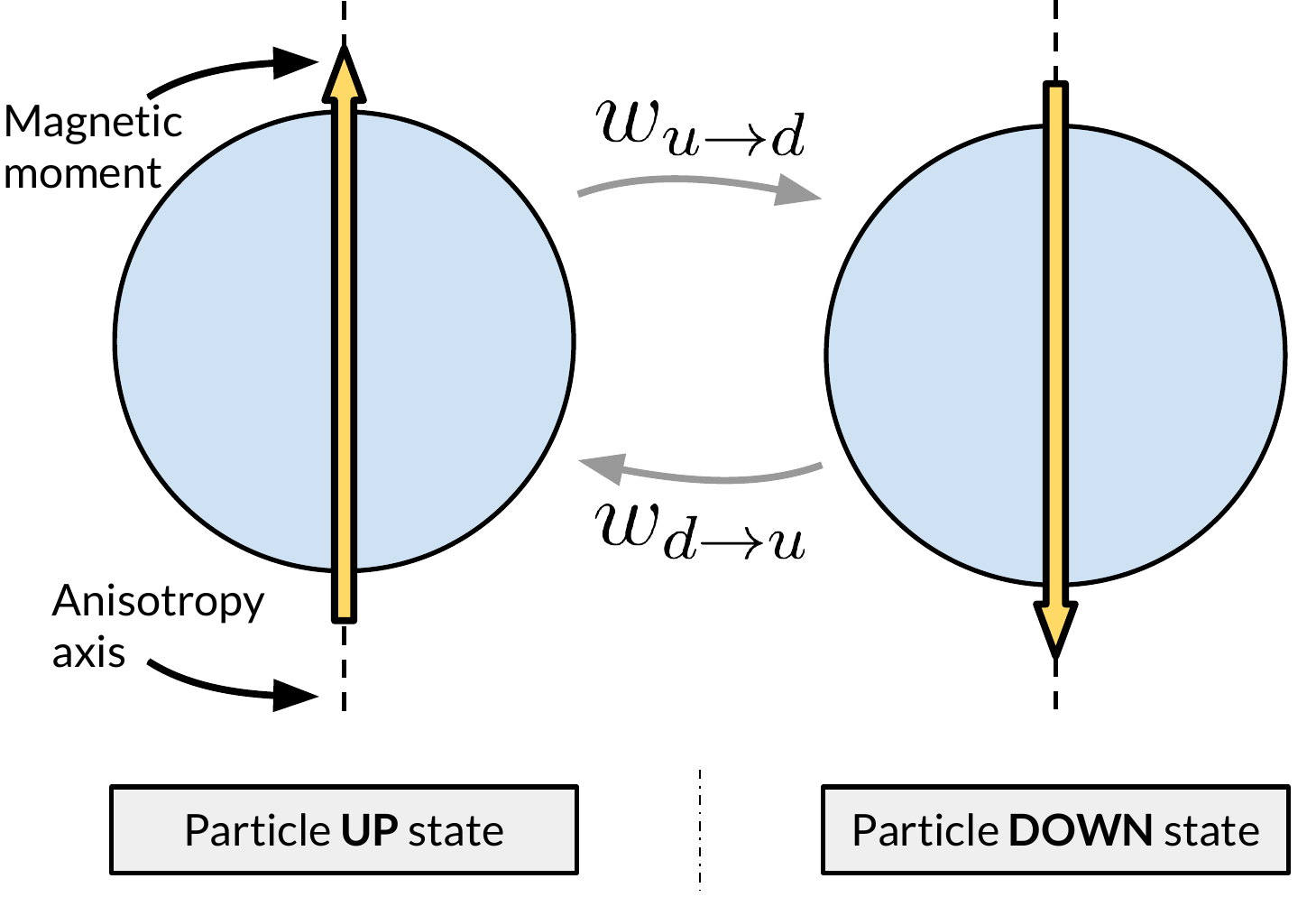}
  \end{subfigure}\hfill%
  \begin{subfigure}[t]{0.45\textwidth}
    \includegraphics[width=\linewidth]{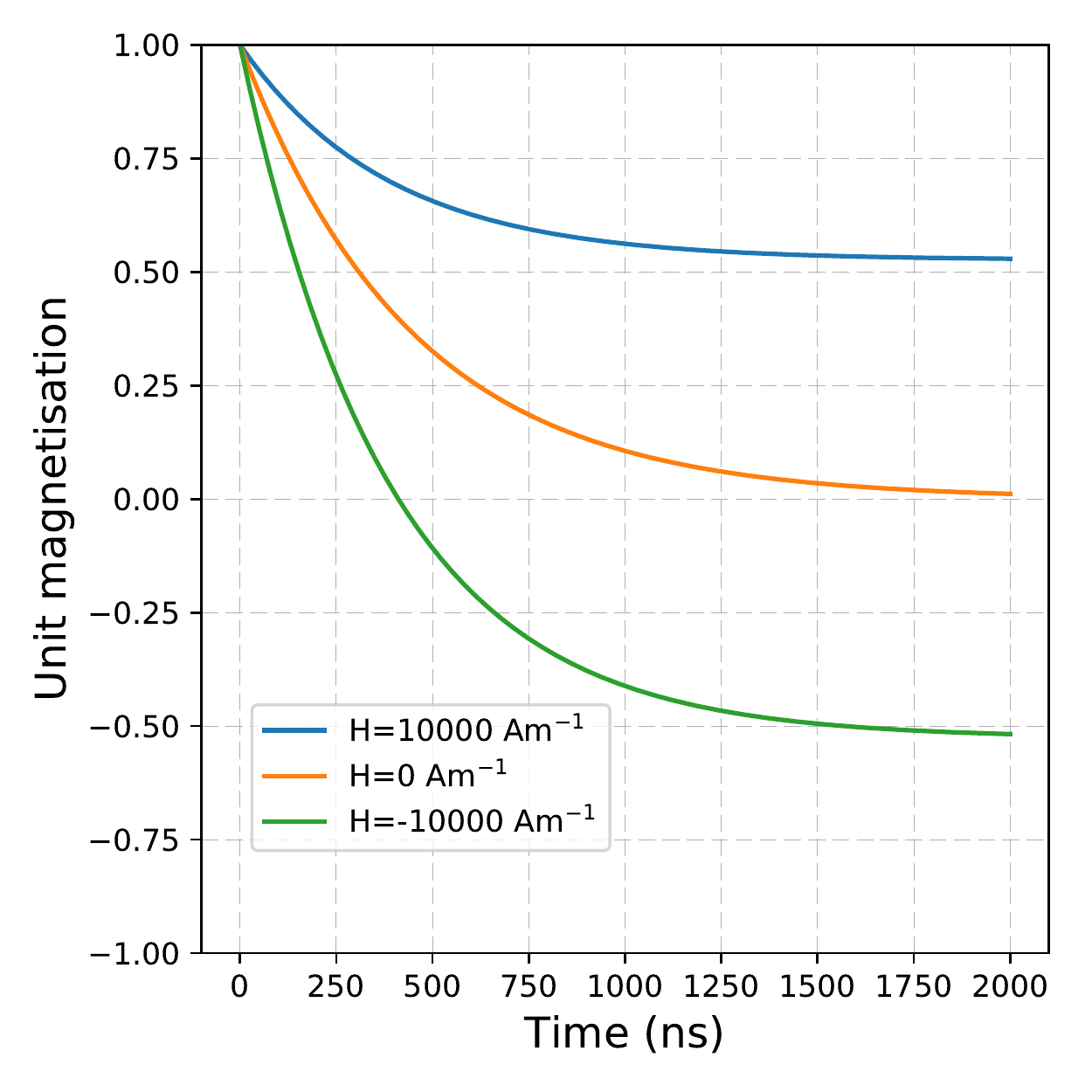}
  \end{subfigure}
  \caption{In the rare-events model, a single nanoparticle may occupy
    one of only two states: \emph{up} or \emph{down}, which each
    correspond to local energy minima around the anisotropy axis
    (left). The dynamics of the system are described by a master
    equation with transition rates $w_{u\to d / d\to u}$ between the
    two states. The solution of the master equation depends on the
    initial condition, particle properties and applied field. In this
    example (right) the particle is initialised up with probability 1
    and allowed to relax into equilibrium (see
    Listing~\ref{listing:dom} for Magpy script). The equilibrium
    magnetisation depends upon the strength of the constant applied
    magnetic field.}
  \label{fig:dom}
\end{figure}

\begin{listing}
  \begin{minted}
  [
  frame=lines,
  framesep=2mm,
  baselinestretch=1.2,
  bgcolor=white,
  fontsize=\footnotesize,
  linenos
  ]
  {python}
import magpy as mp

Hs = [100e2, 0.0, -100e2]
models = [
    mp.DOModel(
        radius=5e-9, anisotropy=5e4, damping=0.01,
        magnetisation=400e3, temperature=300,
        initial_probabilities=[1.0, 0.0], field_amplitude=H)
    for H in Hs
]
results = [
    model.simulate(end_time=2e-6, time_step=1e-10, max_samples=1000)
    for model in models
]
expected_magnetisations = [res.magnetisation() for res in results]
\end{minted}
\caption{Simulating three rare-events models with different applied
  field properties with Magpy (results plotted in Figure
  \ref{fig:dom}). The \texttt{DOModel} object, representing the
  rare-events model, is defined (lines~3-6) by the particle and
  applied field properties. Three identical particles are modelled
  each with a different value of the constant applied field amplitude
  (line~1). Each of the models is simulated (lines~9-12) and the
  expected magnetisation of each model is computed (line~13).}
\label{listing:dom}
\end{listing}

\textbf{Simulating rare-events for single particles}

Magpy provides an alternative, simpler model for simulating
non-interacting, anisotropy-dominated particles. A particle is
considered anisotropy-dominated if the effective field resulting from
anisotropy $k$ is much greater than the thermal fluctuations and the
externally applied field such that $\sigma\qty(1-h)^2\gg1$ (where
$\sigma=KV/\qty(k_BT)$ is termed the reduced energy barrier
height). In these particles, the macrospin remains aligned with one
direction of the anisotropy axis and exhibits long periods of
negligibly small fluctuations around the axis separated by rare-events
in which the macrospin reverses direction. Magpy approximates these
dynamics as a jump process between two discrete states (up and down),
which is described mathematically by the master equation:
\begin{equation}
  \label{eq:master-eq}
  \dv{t}\mqty[p_\mathrm{u}\qty(t) \\ p_\mathrm{d}\qty(t)] = \mqty[-w_{\mathrm{u}\to \mathrm{d}}\qty(t) & w_{\mathrm{d}\to \mathrm{u}}\qty(t) \\
  w_{\mathrm{u}\to \mathrm{d}}\qty(t) & -w_{\mathrm{d}\to \mathrm{u}}\qty(t)] \mqty[p_\mathrm{u}\qty(t) \\ p_\mathrm{d}\qty(t)]
\end{equation}
where the elements of
$\vb{p}\qty(t)=\mqty[p_\mathrm{u}\qty(t),p_\mathrm{d}\qty(t)]^T$ are
the probability that the system is in the up and down state
respectively and
$w_{\mathrm{u}\to \mathrm{d}}\qty(t),w_{\mathrm{d}\to
  \mathrm{u}}\qty(t)$ are the transition rates (units $s^{-1}$)
between the two states. Note that the solution of the master equation
is the time-evolution of the \emph{probability mass function} over the
discrete state space, whereas the solution of the
Landau-Lifshitz-Gilbert equation is the time-evolution of a
\emph{single random trajectory} through the state-space
$\mathbb{R}^3$.

The transition rates are computed in Magpy from the N\'eel-Brown
model, which assumes that the field is applied parallel to the
anisotropy axis direction~\cite{Coffey2012}:
\begin{equation}
  \label{eq:neelbrown}
  w_{\mathrm{u}\to \mathrm{d} / \mathrm{d}\to \mathrm{u}}\qty(t) = \frac{2\gamma\alpha k_BT\sigma^{1.5}\qty(1-h^2\qty(t))}{VM_\mathrm{s}\sqrt{\pi}\qty(1+\alpha^2)}
  \qty(1 \pm h\qty(t))e^{-\sigma\qty(1\pm h\qty(t))^2}
\end{equation}
where $h\qty(t)$ is the reduced applied field magnitude at time $t$
along the anisotropy direction. Equations (\ref{eq:master-eq}) and
(\ref{eq:neelbrown}) are solved numerically from an initial condition
$\vb{p}\qty(t_0)=\mqty[p_\mathrm{u}\qty(t_0),p_\mathrm{d}\qty(t_0)]^T$ at time $t_0$ using an
adaptive step Runge-Kutta solver (RK45) with Cash-Karp parameters
\cite{Press2007}. The total magnetisation at time $t$ for a large
ensemble of particles is computed as
$M\qty(t)=M_\mathrm{s}\qty[p_\mathrm{u}\qty(t)-p_\mathrm{d}\qty(t)]$.

Figure~\ref{fig:dom} shows an example of a single particle simulated
using the Magpy script in Listing \ref{listing:dom}, with a constant
field applied along its anisotropy axis. The initial condition of the
system is $\vb{p}\qty(t_0)=\mqty[1, 0]^T$ and the master equation is
solved numerically. As time evolves, the probability that the particle
flips into the down state $p_\mathrm{d}\qty(t)$ increases and the
expected magnetisation reduces. Eventually, the system reaches an
equilibrium: in zero field the two states are equally likely and the
system has zero magnetisation; for finite applied fields the system
favours one state over the other.

\textbf{Alternative software}

Vinamax\cite{leliaert2015}, implemented in
Golang\footnote{\texttt{https://golang.org/doc} for more information
  on the Go programming language.} and also motivated by the medical
applications of nanoparticles, provides similar functionality to
Magpy. A distinguishing feature of Vinamax is its use of a
multipole-expansion algorithm, which greatly improves the speed of
computing the dipole-dipole interaction forces for large
systems. Magpy computes the interaction field between every pair of
particles (equation (\ref{eq:heff})), an operation with complexity
$\mathcal{O}\qty(n^2)$; the multipole expansion method uses an
approximation, which results in complexity $\mathcal{O}\qty(n\log
n)$. Vinamax implements a broad range of solvers for the LLG dynamics
but currently does not include the fully implicit method.

The purpose of Magpy is not to simulate magnetic systems for which the
macrospin assumption is not justified, such as for larger particles
that exhibit more than a single domain or systems for which
surface-to-surface atomistic interactions are significant. In these
cases, the magnetic moments of the individual atoms must be
modelled. Vampire~\cite{evans2014} is an open-source C++ alternative
to Magpy for atomistic simulation. Vampire reduces the significant
additional computational effort required for simulating individual
atoms by leveraging general purpose graphical processing units
(GPGPUs). Alternatively, if the effects of temperature can be ignored
and the atomistic magnetic moments are closely aligned, the
magnetisation of material can be represented as a continuous function
resulting in a spatial-temporal partial differential equation. This
technique, termed micromagnetics, is implemented in a range of popular
open-source packages: MuMax3~\cite{vansteenkiste2014},
OOMMF~\cite{donahue1999oommf}, fidimag~\cite{fidimag2016},
nmag~\cite{fischbacher2007}.

As discussed, Magpy includes implementations for several numerical
methods to compute approximate solutions to stochastic differential
equations. Currently, the authors are not aware of a reliable
alternative in C++ or Python for the fully implicit method
\cite{milstein2002numerical}. Though there are mature packages for the
solution of ordinary differential equations (e.g. sundials
\cite{hindmarsh2005sundials}) there are few options for stochastic
differential equations. The most mature, SDElab
\cite{gilsing2007sdelab} implemented in Matlab, is no longer under
development and requires proprietary software. A re-implementation of
SDElab using the open-source julia language is currently under
development\footnote{\texttt{https://github.com/tonyshardlow/SDELAB2}
  for development updates on SDELab2.}.

\section*{Implementation and architecture}

Magpy consists of two components. Firstly, a C++ library implements
the core simulation code, which comprises the nanoparticle model and
numerical solvers. The second component is a Python interface to the
C++ library functionality with additional features for setting up
simulations and analysing their results.

The dynamics (Landau-Lifshitz-Gilbert equation), rare-events model,
numerical methods, and the effective field calculations are
implemented in a C++ library. C++ was the preferred programming
language for implementing the computationally-intensive simulation
because of its relatively fast performance and opportunities for
optimisation. The Magpy C++ library is optimised for serial execution;
uses the BLAS and LAPACK libraries; manages memory manually to
minimise allocations and deallocations; and may be compiled with
proprietary Intel compilers for enhanced performance on Intel
architectures. Furthermore, the C++-11 standard contains features that
support a functional programming paradigm (such as closures and
partial application), which were used extensively in Magpy to improve
testability and modularity of code. The entry point to the Magpy
library is through two top-level functions:
\verb!simulation::full_dynamics! for the full model and
\verb!simulation::dom_ensemble_dynamics! for the rare-events
model. Magpy does not provide a graphical user interface, simulations
must be invoked by the user in a C++ program or using the alternative
Python interface.

It was the authors' opinion that scripting in C++ was not sufficiently
usable because of the low-level syntax and the requirement for
compiling scripts, which adds complexity for users. Python, on the
other hand, is high-level, interpreted, and has been gaining
popularity in the computational science community for the design of
user interfaces~\cite{beg2017user,fangohr2016nmag,logg2012dolfin} and
as an easy-to-learn tool~\cite{fangohr2004comparison}. Therefore,
Python was chosen as the preferred language for scripting and
implementing the auxiliary components of Magpy. The interface between
Python and C++ was written using Cython~\cite{behnel2010cython}, which
allowed the C++ library functions to be wrapped as Python functions
and exposed as a Python package, while retaining the performance
benefits of C++. The Python package includes additional features for
building models and plotting the simulation results.

\begin{figure}
  \centering
  \includegraphics[width=\linewidth]{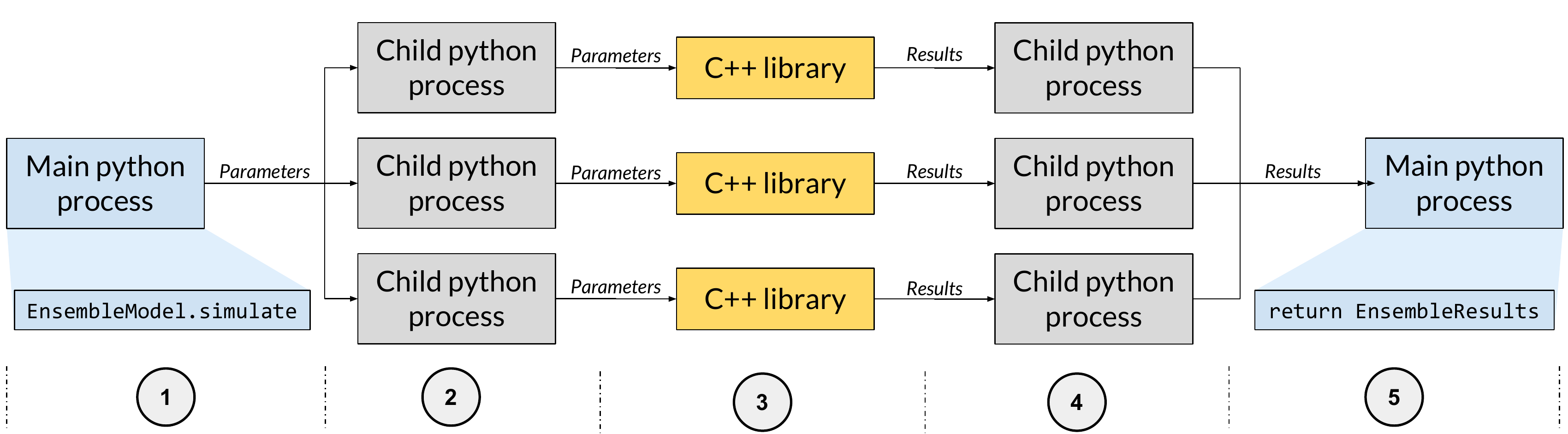}
  \caption{The flow of data through an ensemble simulation in
    Magpy. (1)~The user instantiates an \texttt{EnsembleModel} object
    and calls the \texttt{EnsembleModel.simulate} function, specifying
    the number of CPU cores to utilise. (2)~The main Python process
    spawns a new individual process for each model in the
    ensemble. (3)~The individual processes each call the Magpy C++
    library using their respective model parameters. (4)~The results
    from the C++ simulation are returned to the individual python
    process. (5)~When all the processes have finished, the results are
    collected on the main process to be analysed and plotted.}
  \label{fig:process}
\end{figure}

The typical workflow for a Magpy experiment consists of running
multiple simulations of the same model in order to generate a
distribution of possible trajectories, as in
Figure~\ref{fig:sim-example}. This motivates an embarrassingly
parallel strategy in which each simulation executes concurrently on a
single process, since no communication is required between the
independent runs. Parallelism is implemented in Python using
joblib\footnote{\texttt{https://pythonhosted.org/joblib/} for the
  joblib documentation.}. A minimum
example of how joblib is used to execute tasks in parallel is shown in
Listing~\ref{listing:parallel}. In Magpy, the user creates an ensemble
of models (the \texttt{EnsembleModel} object in
Listing~\ref{listing:threechain} lines~3-18) and begins the simulation
(\texttt{EnsembleModel.simulate} lines~19-20) utilising the requested
number of cores (\texttt{n\_jobs}). For each model in the ensemble,
Magpy creates a new independent python process containing a copy of
the model object. Each process then simulates its respective model by
calling functions in the C++ library with the model parameters. As
many as \texttt{n\_jobs} simulations may execute concurrently. Once
each simulation finishes, the results are returned from the C++
library to the individual python process. Once all processes have
completed, the results are gathered back into the python process with
which the user was originally interacting. This architecture is
displayed graphically in Figure~\ref{fig:process}.

\begin{listing}
  \begin{minted}[
  frame=lines,
  framesep=2mm,
  baselinestretch=1.2,
  bgcolor=white,
  fontsize=\footnotesize,
  linenos
  ]
  {python}
from joblib import Parallel, delayed
import time

def slow_double(x):
    time.sleep(1) # 1 second sleep
    y = 2*x
    return y

xs = [2, 6, 12, 24, 40, 72, 126, 240]

# Serial computation takes approximately 8s
ys_serial = [slow_double(x) for x in xs]
print(ys_serial)
#> [4, 12, 24, 48, 80, 144, 252, 480]

# Embarrassingly parallel computation takes approximately 2s
ys_parallel = Parallel(n_jobs=4)(delayed(slow_double)(x) for x in xs)
print(ys_parallel)
#> [4, 12, 24, 48, 80, 144, 252, 480]
  \end{minted}
  \caption{A minimal example of an embarrassingly parallel computation
    with \texttt{joblib}. The function \texttt{slow\_double} (line 4)
    doubles a single number and takes approximately one second. The
    objective is to evaluate the function with eight different
    arguments (line 9). This is achieved in serial with a \texttt{for}
    loop (line 12) by evaluating the function for each argument in the
    list singly, taking approximately eight seconds. However, this
    problem is embarrassingly parallel because all evaluations of
    \texttt{slow\_double} may occur concurrently since each function
    call only depends on its initial argument. Using joblib, the eight
    function calls are evaluated on four processes as shown (line 17)
    by setting \texttt{n\_jobs=4}. Two evaluations are distributed to
    each of the four processes, which execute concurrently, taking
    approximately two seconds.}
  \label{listing:parallel}
\end{listing}

\section*{Quality control}
Magpy has been tested to increase confidence in the correctness of the
implementation, mathematics, and physics. The lowest level of tests,
unit tests, assert that individual functions return the correct answer
given a set of fixed arguments. The unit tests are designed to catch
bugs during development and test the installation of the
software. Continuous integration, using
CircleCI\footnote{https://circleci.com/ for more information.},
ensures that tests are automatically executed before changes are
committed to the existing code repository on Github. The unit tests
are implemented using
GoogleTest\footnote{\texttt{https://github.com/google/googletest} for
  the GoogleTest repository.} for C++ functions and
pytest\footnote{\texttt{https://docs.pytest.org/en/latest/} for the
  pytest documentation.} for Python functions.

Numerical tests are necessary to confirm the stability and robustness
of the numerical methods. Magpy includes scripts to evaluate the
empirical convergence rates of the SDE solvers and compares them with
analytic solutions \cite{Kloeden1992b,milstein2002numerical}. The
numerical tests should be used during the development of new or
existing solvers.

Finally, Magpy includes a series of Jupyter notebooks\footnote{Magpy
  documentation and examples are hosted at
  \texttt{http://magpy.readthedocs.io}.} that present tutorials and
examples, including comparisons of simulation results with theoretical
solutions from alternative models in physics. These comparisons assert
that the simulations, under the appropriate assumptions, correctly
approximate the magnetic nanoparticle dynamics. The fundamental
benefit of Jupyter notebooks is that they contain documentation,
mathematics, figures, and executable code in a single format. However,
this introduces additional maintenance, the example code in the
notebooks must be updated when the interfaces or structure of the
program changes. Therefore, we used the nbval
tool\footnote{\texttt{https://github.com/computationalmodelling/nbval}
  for the nbval repository.} as part of our testing practices to
validate the consistency of the notebooks. The validation process
asserts that the code examples run without error and that their result
is consistent with the most recent documentation. Developers are
expected to validate all existing notebooks before committing changes
to the code base to ensure that they remain a relevant and executable
form of documentation for users.

\section*{(2) Availability}
\vspace{0.5cm}
\section*{Operating system}

Available for all Linux based systems. Tested on Ubuntu 16.04 and
RedHat 6.3.

\section*{Programming language}

Python version 3.5 and a C++11/14 compatible compiler (e.g. G++-4.9
and above)

\section*{Additional system requirements}

A single modern processor and 1GB of RAM is sufficient for basic
models. 10MB of disk space is required for the Magpy source code, and
a total of 35MB for the compiled libraries and interface.

\section*{Dependencies}

g++-4.9, python3, openblas, setuptools, cython, numpy, matplotlib,
toolz, joblib, scipy, transforms3d, pytest, nbval

\section*{List of contributors}

Oliver W. Laslett, Jonathon Waters, Hans Fangohr, Ondrej Hovorka

\section*{Software location:}

{\bf Archive}
\begin{description}[noitemsep,topsep=0pt]
	\item[Name:] Zenodo
	\item[Persistent identifier:] 10.5281/zenodo.1124942
	\item[Licence:] 3-Clause BSD
	\item[Publisher:]  Oliver Laslett
	\item[Version published:] 1.1
	\item[Date published:] 14/01/18
\end{description}

{\bf Code repository}

\begin{description}[noitemsep,topsep=0pt]
	\item[Name:] Github
	\item[Persistent identifier:] \texttt{https://github.com/owlas/magpy/tree/v1.1}
	\item[Licence:] 3-Clause BSD
	\item[Date published:] 14/01/18
\end{description}

\section*{Language}

English

\section*{(3) Reuse potential}

Magpy was primarily designed for simulating the magnetic dynamics of
nanosized particles. The simulation results may be used to compute
heat dissipation, relaxation rates, and equilibrium states allowing
the software to help predict, explain, or otherwise augment
traditional experiments in the laboratory or clinical
settings. However, using numerical simulation also allows the
exploration of a range of material geometries and properties without
expensive equipment or physical limits. Magpy was designed to be
accessible to experts and non-experts through the extensive
documentation and included examples.

In addition to its uses in physics, the implementation of the
numerical solvers for stochastic differential equations may be useful
beyond the original purpose of Magpy. The Landau-Lifshitz-Gilbert
equation belongs to a class of equations (multi-dimensional,
nonlinear, stochastic, non-commutative, stiff) that are challenging to
solve numerically. Magpy could also be used for teaching concepts in
magnetism as the Python interface will likely be familiar to new
students in physics.

A number of additional features remain that have yet to be implemented
in Magpy. In particular, the use of a multiplole expansion method
(inspired by Vinamax) would reduce the time required to compute the
interaction fields. It would also be possible to extend Magpy to
simulate atomistic-level dynamics by decomposing each macrospin into a
lattice of atomistic moments and including the exchange interaction
term to the effective field. The rare-events model currently supports
a single particle with the field applied along its anisotropy axis.
Allowing arbitrary applied field directions as well as dipole-dipole
interactions between multiple particles would greatly increase the
potential applications of the model.

Contributions are welcomed and encouraged in the form of feature
requests, bug reports, and suggestions for new algorithms. All
communication should be directed to the github repository in order to
keep a publicly available record of issues, which may be addressed
by the members of the community. Currently, active support is limited
to the lead developer (Oliver Laslett) but we hope that improved
support will result from a growing user base.

\section*{Acknowledgements}

Many thanks to all members of the Computational Engineering and Design
group for enlightening discussions on research software design. A
special thank you to Thomas Kluyver for his unfathomable depth of
knowledge of the Python ecosystem.

\section*{Funding statement}

The authors gratefully acknowledge financial support from the EPSRC
doctoral training centre grant (EP/G03690X/1).

\section*{Competing interests}

The authors declare that they have no competing interests.

\bibliographystyle{unsrtnat}
\bibliography{../../../Literature/BIBLIOGRAPHY}

\vspace{2cm}

\rule{\textwidth}{1pt}

{ \bf Copyright Notice} \\
Authors who publish with this journal agree to the following terms: \\

Authors retain copyright and grant the journal right of first publication with the work simultaneously licensed under a  \href{http://creativecommons.org/licenses/by/3.0/}{Creative Commons Attribution License} that allows others to share the work with an acknowledgement of the work's authorship and initial publication in this journal. \\

Authors are able to enter into separate, additional contractual arrangements for the non-exclusive distribution of the journal's published version of the work (e.g., post it to an institutional repository or publish it in a book), with an acknowledgement of its initial publication in this journal. \\

By submitting this paper you agree to the terms of this Copyright
Notice, which will apply to this submission if and when it is
published by this journal.

\end{document}